\documentclass[fleqn,usenatbib]{mnras}

\usepackage{newtxtext,newtxmath}
\usepackage{xspace}

\usepackage[T1]{fontenc}
\usepackage{natbib}
\DeclareRobustCommand{\VAN}[3]{#2}
\let\VANthebibliography\thebibliography
\def\thebibliography{\DeclareRobustCommand{\VAN}[3]{##3}\VANthebibliography}

\usepackage{graphicx}	
\usepackage{amsmath}	

\title[]{ Intranight variability of UV emission from powerful blazars}
\author[$Chand$ et al.]{{ \large Krishan Chand$^{1,2}$\thanks{E-mail: krishan@aries.res.in(KC)},
Gopal-Krishna$^{3}$, Amitesh Omar$^{1}$, Hum Chand$^{4,1}$, Sapna Mishra$^{5}$, P. S. Bisht${^6}$, and S. Britzen$^{7}$}\\\\
$^{1}$Aryabhatta Research Institute of Observational Sciences (ARIES), Manora Peak, Nainital $-$ 263 002, India\\
  $^{2}$Kumaun University, Nainital $-$ 263 002, India\\
  $^{3}$UM-DAE Centre for Excellence in Basic Sciences, Vidyanagari, Mumbai-400098, India\\
  $^{4}$Central University of Himachal Pradesh, Temporary Academic Block, Shahpur,
  Kangra, Himachal Pradesh-176206, India\\
  $^{5}$Inter-University Centre for Astronomy and Astrophysics (IUCAA), Postbag 4, Ganeshkhind, Pune 411 007, India\\
   $^{6}$Department of Physics, Soban Singh Jeena University, Almora - 263601\\
  $^{7}$  Max-Planck-Institut f. Radioastronomie, Auf dem Huegel 69, D-53121 Bonn, Germany\\
}

\begin{document}
\date{Accepted ---. Received ---; in original form ---}

\pagerange{\pageref{firstpage}--\pageref{lastpage}} \pubyear{2021}

\maketitle

\label{firstpage}
\begin{abstract}
  We report the first study to characterise intranight variability of the blazar class from the perspective of (rest-frame) UV emission. For this, we carried out intranight optical monitoring of 14 flat-spectrum radio quasars (FSRQs) located at high redshifts (1.5 < $z$ <  3.7), in 42 sessions of median duration $\sim$ 5.4 hr. These sources were grouped into two samples distinguished by published fractional optical polarisation:  (i) nine low-polarisation sources with $p_{opt} < 3\%$ and  (ii)  five high-polarisation sources. Unexpectedly, a high duty cycle (DC $\sim$ 30\%) is found for intranight variability (with amplitude $\psi > 3\%$) of the low-polarisation sources. This DC is a few times higher than that reported for low-polarisation FSRQs located at moderate redshifts ($z$ $\sim$ 0.7) and hence typically monitored in the rest-frame blue-optical. Further, we found no evidence for an increased intranight variability of UV emission with polarisation, in contrast to the strong correlation found for intranight variability of optical emission. We briefly discuss this in the context of an existing scenario which posits that the nonthermal UV emission of blazars arises from a relativistic particle population different from that radiating up to near-infrared/optical frequencies.
 \end{abstract}
\section{Introduction}
\label{introduction}
According to the current paradigm, blazars are Active Galactic Nuclei (AGN) whose observed radiation is predominantly nonthermal and arises from a relativistic jet roughly pointed and hence beamed towards the observer \citep{Blandford1978PhyS...17..265B,Moore1984ApJ...279..465M,Antonucci1993ARA&A..31..473A,Urry1995PASP..107..803U}. Well known manifestations of such beaming include a flat radio spectrum (i.e., a dominant radio core) and rapid variability of continuum and polarised emission. The flat-spectrum radio quasars (FSRQs) are identified as a quasar subset of blazars with higher power \citep[e.g.,][]{Stockman1984ApJ...279..485S,Wills1992ApJ...398..454W,Angelakis2016MNRAS.463.3365A}. 
Their jets frequently exhibit superluminal motion and flux variability from radio to gamma-rays, on time-scales as short as minutes. The variability studies have richly contributed to understanding the nature of their central engines and jets
\citep[e.g.,][]{Wagner1995ARA&A..33..163W,Aharonian2017ApJ...841...61A,Gopal2018BSRSL..87..281G}. However, information on the intranight variability is conspicuously lacking for the ultra-violet (UV) emission from quasars. This is particularly unsatisfactory in view of the hint emerging from detailed spectral measurements of a few quasar jets, that a very substantial, if not dominant contribution to their synchrotron UV radiation may arise from a relativistic particle population distinct from the one responsible for their radiation up to near-IR and optical frequencies. This inference is based on detailed spectra of the knots in the kiloparsec-scale radio jets of some quasars, which are found to exhibit an {\it upturn} towards the UV and then connecting smoothly to X-ray data points \citep{Uchiyama2006ApJ...648..910U,Uchiyama2007ApJ...661..719U,Jester2007MNRAS.380..828J,Meyer2017ApJ...835L..35M}. The synchrotron origin of this high-energy component is confirmed by the observed high polarisation of the UV emission \citep{Cara2013ApJ...773..186C}. The importance of finding independent manifestations of this putative secondary population of relativistic particles has been pointed out in \citet{Gopal-krishna2019BSRSL..88..132G}.\par
A key reason for the scarcity of information about rapid rest-frame UV variability of AGN is that intranight monitoring campaigns are prohibitively time expensive for the (space-borne) UV telescopes \citep[e.g.,][]{Smith2018ApJ...857..141S}. In recent years, this issue has begun to get addressed by taking advantage of the UV data acquired with {\it GALEX} for large AGN samples. However, such studies have only covered day-like, or longer time scales and are focused on optically selected quasars \citep[e.g.,][]{Punsly2016ApJ...830..104P,Xin2020MNRAS.495.1403X}, thus probing the accretion disk rather than the jet.
In order to probe rapid UV variability of blazars, a practical approach adopted here is to perform intranight {\it optical} monitoring of blazars located at sufficiently high redshifts, so that the monitored optical radiation is their rest-frame UV emission. A similar strategy had earlier been followed for investigating UV variability of {\it optically selected} quasars on month/year-like time scales \citep[e.g.,][]{Wilhite2005ApJ...633..638W}. The present study encompasses two prominent subclasses of blazars, distinguished by low and high fractional polarisation measured in the optical, with the division taken at $p_{opt} = 3\%$ \citep{Moore1984ApJ...279..465M}. It may be recalled that for moderately distant blazars, a strong positive correlation between intra-night optical variability (INOV) and $p_{opt}$ is found, inspite of the fact that the two sets of observations had been made a decade apart \citep[][and references therein]{Goyal2012AA...544A..37G}. The longer-term optical variability is also known to be stronger for sources with higher $p_{opt}$ \citep[e.g.,][]{Angelakis2016MNRAS.463.3365A}. The present study examines intranight variability of the blazar class, for the first time, from the perspective of the rest-frame UV emission.

\section{Samples of high-$z$ blazars} \label{The two blazar samples}
Our first sample of high-$z$ blazars for intranight optical monitoring is derived from the 5th edition of ROMA blazar catalogue (ROMA-BZCAT, \citep{Massaro2015Ap&SS.357...75M}) which lists 3561 sources classified either as confirmed blazars, or exhibiting characteristics close to blazars.
Out of these, 420 blazars have $z > 2$. We selected brighter sources with $m_R \leq$ 17 mag since they are suitable for monitoring with the metre-class telescopes. This left us with 18 blazars. Of these, 7 sources at negative declinations were discarded, being too far south for monitoring with the telescopes at ARIES, Nainital. Another source, J185027.59+282513.18 was discarded since its long intranight monitoring is not possible as its nightly transit occurs during the monsoon period at Nainital. The last source to be excluded is J161942.38+525613.41, because of non-availability of polarisation information. 
Thus, the final sample has 9 low-polarisation (median $p_{opt}$ $\sim 0.8\%$) flat-spectrum radio quasars ($LP_{FSRQs})$ with $z > 2$ (sample-1, Table (1a), see also online Table 1).\par

Since our objective is also to investigate a possible dependence of intranight variability on polarisation for high-$z$ blazars, we also constituted a high-polarisation counterpart to the first sample and this sample-2 is comprised of $HP_{FSRQs}$. Given the rarity of such sources at high redshifts, the selection criteria for this second sample were slightly relaxed (see below). Also, 
since only one source in the ROMA-BZCAT was found to meet those selection criteria, we constituted the sample-2 
(Table (1b)) 
from the optopolarimetric survey RoboPol \citep{Angelakis2016MNRAS.463.3365A} instead of using ROMA-BZCAT. Out of the total 158 $\gamma-ray$ detected blazars observed in RoboPol, we first selected those listed with $z > 1.5$ and $p_{opt} >  3\%$ and then applied additional filters considering the observability with the available telescope: (a) R.A. between 23h and 17h \& dec. $>$ 0 deg;  and (b) $m_R < 17.5$. One qualifying source, J1314$+$2348 had to be discarded because its redshift turned out to be highly uncertain (e.g., $z \sim 0.150$, \citealp{Makarov2019ApJ...873..132M}), due to lack of features in its spectrum. This is not entirely unexpected since at very high luminosities, optical spectrum of even a quasar can appear featureless \citep[e.g.,][]{Wills1983ApJ...274...62W}. But we note that high-polarisation quasars are known to exhibit strong INOV, regardless of visibility of emission lines, or lack of them in the optical spectrum \citep[e.g.,][]{GoyalA2013MNRAS.435.1300G}. Thus, our sample-2 consists of 5 sources, termed $HP_{FSRQs}$, all of which have a high redshift ($z > 1.5$) and a flat radio spectrum which signifies a dominance by a relativistically-beamed jet (Table (1b)).
It may be reiterated that the combination of the main selection criteria employed here for both our samples, namely a large redshift and high apparent optical brightness, implies that these samples are populated by blazars belonging to the highest luminosity tail of blazars, a point further discussed in Sect.~\ref{Discussion}.
\begin{table}
  \renewcommand\thetable{}
  \label{tab:Table}
 
  \resizebox{1.1\columnwidth}{!}{%
    \begin{tabular}{cccccccccccc}
  \hline\\
  \multicolumn{8}{c}{\large {\bf Table} (1a): Sample-1 containing 9 bright $LP_{FSRQs}$ at $z > 2.0$.}\\\\
  \hline\\
\multicolumn{1}{c}{SDSS name}  & \multicolumn{1}{c}{R.A. (J2000)} &
\multicolumn{1}{c}{Dec (J2000)} & \multicolumn{1}{c}{$z$}
 &
\multicolumn{1}{c} {$R_{mag}$}& \multicolumn{1}{c}{$\alpha_{radio}$} &\multicolumn{1}{c}{$p_{opt}$} & Ref. & \\ 
& hh:mm:ss   &$^\circ$: $^\prime$: $^{\prime\prime}$   & &     &   &(\%)  &codes&\\
(1)        &(2)         &(3)                      &(4)        &(5)     &(6)&(7)&(8)&\\\\
\hline\\

 J001708.45+813508.19 &  00:17:08.45 & +81:35:08.19 &  3.387 & 15.9 &   -0.18  &$1.10\pm0.50$    &a&\\
     &              &             &         &      &          & 0.60\xspace\xspace\xspace           &b&\\
     &              &             &         &      &           &$0.99\pm0.27$  &c&  \\
 J015734.96+744243.20 &  01:57:34.96 & +74:42:43.20 &  2.341 & 16.7 &-0.10  & $0.49\pm1.52$ &c&\\
    &              &             &           &     &           & $1.10\pm0.30$  &d&\\
    &              &             &         &       &           & 1.72\xspace\xspace\xspace          &e&\\
 J042408.56+020424.99  &  04:24:08.56 & +02:04:24.99 &  2.056 & 16.4 &-0.38  & $0.95\pm0.29$ & c&\\
   &              &             &           &      &           & 0.31\xspace\xspace\xspace          &e &\\
 J064204.26+675835.61 &  06:42:04.26 & +67:58:35.61 &  3.180 & 16.1  &\xspace0.73   & 0.70\xspace\xspace\xspace         & b&\\
 J084124.35+705342.28 &  08:41:24.35 & +70:53:42.28 &  2.218 & 16.8 &-0.38  &$2.00\pm0.60$    & a&\\
   &              &             &           &      &          &$0.77\pm0.21$  & c&\\
   &              &             &           &      &           &1.10\xspace\xspace\xspace           & f&\\
 &              &             &           &      &           &$1.10\pm0.50$  & g&\\
 J122824.96+312837.59 &  12:28:24.96 & +31:28:37.59 &  2.195 & 15.8 &\xspace0.03    &$0.16\pm0.24$ & h&\\
   &              &             &           &      &          &0.16\xspace\xspace\xspace           &e&\\
 J133335.77+164904.00 &  13:33:35.77 & +16:49:04.00 &  2.084 & 16.1 &\xspace0.36   &$0.27\pm0.26$ & c&\\
   &              &             &           &      &            & 0.37\xspace\xspace\xspace        &e &\\
 J142438.10+225600.99 &  14:24:38.10 & +22:56:00.99 &  3.620 & 15.2 &\xspace0.56    &$0.40\pm0.12$ & i&\\
 J143645.79+633637.90 &  14:36:45.79 & +63:36:37.90 &  2.068 & 17.0 &-0.18   &$1.20\pm0.40$     &a &\\
    &              &             &           &     &             & 0.80\xspace\xspace\xspace            &b &\\
    &              &             &           &     &             & $0.50\pm0.29$  &c &    \\
    &              &             &           &     &             & $0.26\pm0.52$  &e &  \\\\
 \hline\\
 \multicolumn{8}{l}{ Notes. Col. 2,3: coordinates; Col. 4: emission redshift;}\\
 \multicolumn{8}{l}{Col. 5: R-band magnitude from ROMABZ; Col. 6: radio spectral index ($f_{\nu} \propto \nu^{\alpha}$);}\\
 \multicolumn{8}{l}{Col. 7: fractional optical polarization; Col. 8: reference code(s).} \\
 \multicolumn{8}{l}{ References: (a) \citet{Pavlidou2014MNRAS.442.1693P}; (b) \citet{Kovalev2020MNRAS.493L..54K}; (c) \citet{Wills2011ApJS..194...19W};}\\
\multicolumn{8}{l}{ (d) \citet{Tornia2008AA...482..483T}; (e) \citet{Wills1992ApJ...398..454W}; (f) \citet{Joshi2007MNRAS.380..162J}; }\\
\multicolumn{8}{l}{ (g) \citet{Hutse2005AA...441..915H}; (h) \citet{Hutse2001AA...367..381H}; (i) \citet{Afanas2014ARep...58..725A}.}\\\\
\hline\hline\\
\multicolumn{8}{c}{\large {\bf Table} (1b): Sample-2 containing 5 bright $HP_{FSRQs}$ at $z > 1.5$.}\\\\
\hline\\
SDSS name & RA (J2000) & Dec (J2000) &$z$  &$R_{mag}$&$\alpha_{radio}$ &$p_{opt}$ & Ref.  \\
& hh:mm:ss   &$^\circ$: $^\prime$: $^{\prime\prime}$   & &     &   &(\%)  &codes\\\\
\hline\\
J011452.77+132537.53&01:14:52.77  & +13:25:37.53 &  2.025 & 17.2 &-0.07& $6.6\pm0.1$    & a&\\
  &             &              &        &       &     &  $7.2\pm1.2$   & b&\\
 &             &              &        &       &     &  $9.1\pm0.6$   & e&\\
J105430.60+221055.00& 10:54:30.60  & +22:10:55.00 &  2.055 & 17.1&-0.13& $6.7\pm1.5$    & a&\\
 &             &              &        &       &     &  $7.3\pm1.3$ & c&\\

J124510.00+570954.37	& 12:45:10.00  & +57:09:54.37 &  1.545 & 17.4r&\xspace0.14 & $14.4\pm1.5$ &a& \\
&             &              &        &       &     &  $16.9\pm1.3$ & c&\\

J163515.49+380804.50& 16:35:15.49  & +38:08:04.50 &  1.813 & 17.3&\xspace0.13 & $9.9\pm0.1$   & a&\\
&             &              &        &       &     &  $7.0\pm0.5$ & d&\\

J164924.99+523514.99& 16:49:24.99  & +52:35:14.99 &  2.055 & 17.0&-0.05 & $4.5\pm0.1$  & a&\\
 &             &              &        &       &     &  $9.0\pm0.4$ & c&\\\\
\hline\\
   \multicolumn{8}{l}{ References: (a) \citet{Angelakis2016MNRAS.463.3365A}; (b) \citet{Hovatta2016AA...596A..78H}; (c) \citet{Pavlidou2014MNRAS.442.1693P};}\\
   \multicolumn{8}{l}{ (d) \citet{Lister2000ApJ...541...66L}; (e) \citet{Kuegler2014AA...569A..95K}.}\\
  \end{tabular}
  }
  
\end{table}

 \section{Photometric Monitoring}
 \label{Photo}
 Intranight monitoring of the 9 high-$z$ $LP_{FSRQs}$ (sample-1) and the 5 high-$z$ $HP_{FSRQs}$ (sample-2) was carried out in    Johnson-Cousins R or SDSS r-band in a total of 27 and 15 sessions of median durations 5.3 hr and 5.8 hr, respectively (3 sessions per source). 
  The telescopes used are the 1.3-m Devasthal Fast Optical Telescope (DFOT; \citet{Sagar2011CSci..101.1020S}, 36 sessions), the 1.04-m Sampuranand Telescope (ST; \citet{Sagar1999CSci...77..643S}, 4 sessions) and the 3.6-m Devasthal Optical Telescope (DOT; \citet{Kumar2018BSRSL..87...29K}, 2 sessions). 
The observations were carried out using the 4k $\times$ 4k (0.23$^{\prime\prime}$/pixel) liquid nitrogen cooled CCD on the 1.04-m ST with a binning of 4 $\times$ 4, the 2k $\times$ 2k (0.53$^{\prime\prime}$/pixel) Peltier-cooled CCD on the 1.3-m DFOT without binning and the ARIES Devasthal-Faint Object Spectrograph and Camera (ADFOSC) \citep{Amitesh2019arXiv190205857O} on the 3.6-m DOT. The ADFOSC ($\sim$ 0.2$^{\prime\prime}$/pixel) is equipped with a deep depletion 4k $\times$ 4k CCD, cooled to $-120^{\circ}$C using a closed-cycle cryo cooling thermal engine. A binning of 4 $\times$ 4 was used for the ADFOSC observations.
\subsection{Data Reduction} \label{Data Reduction}
The standard procedure was followed for pre-processing and cleaning of the CCD frames. The cleaned frames were aligned using the {\small PYTHON} package {\small ALIPY}. The instrumental magnitude of the blazar and the 3 steady appearing comparison stars contained in each CCD frame, were determined through aperture photometry \citep[see,][]{Stetson1987PASP...99..191S,1992ASPC...25..297S}, using the Dominion Astronomical Observatory Photometry II (DAOPHOT II) package. The aperture radius, a crucial parameter for photometry, was set equal to two times the point spread function (PSF), which was estimated by averaging the FWHMs of the Gaussians fitted to the brightness profiles of 5 moderately bright stars within the frames. The PSF variations during a session are plotted in the lowest panel in each of the online Figs. 1-8.
However, for the 3 sessions devoted to J0157$+$7442, the aperture radius was set equal to the PSF, in order to minimise the expected confusion from the relatively bright point-like object seen at a separation of 6.5 arcsec. For each session, we then derived differential light curves (DLCs) for all pairs involving the target blazar and the chosen 3 comparison stars 
(online Figs. 1-8; online Tables 2 and 3).
\section{Statistical Analysis}
\label{stat}
The presence of INOV in the DLCs was checked by applying the $F_{\eta}$ test \citep{Diego2010AJ....139.1269D}, following the basic procedure described in \citet{Sapna2019MNRAS.489L..42M}. Among the 3 comparison stars initially chosen for a given session, we identified the steadier two by inspecting the star-star DLCs and the $F_{\eta}$ test was applied to only the 3 DLCs involving those two comparison stars and the target blazar. The selected two comparison stars for each session are shown within parentheses in column 5 of the online Table 4 and the labels of the 3 DLCs involving them are placed within parentheses in the online Figs. 1-8, on the right side. Recall that in several independent studies, it has been shown that the photometric errors returned by DAOPHOT are too small \citep{Gopal-Krishna1995MNRAS.274..701G,Stalin2004JApA...25....1S,Bachev2005MNRAS.358..774B} by a factor $\eta$ for which an accurate estimate of {$1.54\pm0.05$} has been made in an exhaustive analysis employing a large set of 262 DLCs of pairs of steady comparison stars, monitored in 262 intranight sessions of minimum 3 hour duration each, targeted at quasars/blazars \citep{Goyal2013JApA...34..273G}. In the present study we have adopted the same value of $\eta$ ( = 1.54) because of the high precision of this estimate. In addition, this choice of $\eta$ value ensures consistency as we have made a direct comparison of our INOV results with those derived for several prominent subclasses of blazars/quasars (including those studied here), by \citet{GoyalA2013MNRAS.435.1300G} in an extensive work where the same $F_{\eta}$ statistical test was applied and $\eta$ was also taken to be 1.54. \par

The $F$-values for the selected two blazar DLCs for a session are:
\begin{equation} 
\label{eq.ftest2}
F_{1}^{\eta} = \frac{Var(q-s1)}
{ \eta^2 \sum_\mathbf{i=1}^{N}\sigma^2_{i,err}(q-s1)/N},  \\
\hspace{0.1cm} F_{2}^{\eta} = \frac{Var(q-s2)}
{ \eta^2 \sum_\mathbf{i=1}^{N}\sigma^2_{i,err}(q-s2)/N}  \\
\end{equation}\\
where $Var(q-s1)$ and $Var(q-s2)$ are the variances of the DLCs of the target blazar, relative to the selected two comparison stars,
and $\sigma_{i,err}(q-s1)$ and $\sigma_{i,err}(q-s2)$ represent the rms error returned by DAOPHOT on the $i^{th}$ data point in the DLCs of the target blazar, relative to the two comparison stars.
N is number of data points in the DLCs and the scaling factor  ${\eta= 1.54}$, as mentioned above.
The online Table 4 lists N and the computed values of $F_{1}^{\eta}$ and $F_{2}^{\eta}$ for the selected two DLCs of the target blazar for each session.\par
The critical values of $F $ ($= F_{c}^{\alpha}$) for $\alpha = $ 0.05, 0.01 correspond to confidence levels of 95\% and 99\%, respectively. For each session, these two computed values are listed in columns 6 $\&$ 7 in the online Table 4 and they are compared with the $F-$values computed for the two DLCs of the blazar using Eq. (1), namely, $F_{1,2}^{\eta}$ (Column 5 in the online Table 4). If the computed $F-$value for a DLC of the target blazar exceeds the critical value $F_{c}$ for that session, the null hypothesis (i.e., no variability) is discarded. For a computed $F-$value $\ge$ $F_{c}$(0.99), the DLC of the target blazar is classified as `variable' (V). The designation is `probable variable' (PV) if the computed $F-$value falls between $F_{c}$(0.95) and $F_{c}$(0.99), and `non-variable'  (NV) if the $F-$value is less than $F_{c}$(0.95). Note that the designation for a given session, as given in the column 9 of the online Table 4, is `V' only if both DLCs of the target blazar belong to the `V'  category and `NV' if even one of the two DLCs is of the `NV' type. The remaining sessions have been designated `probable variable' (PV). The column 10 of the online Table 4 lists for each session the `Photometric Noise Parameter' (PNP) = {$\sqrt { \eta^2\langle \sigma^2_{i,err} \rangle}$ }, where ${\eta=1.54}$, as mentioned above.

The variability amplitude ($\psi$) for a DLC is defined as \citep{Heidt1996A&A...305...42H}:\\
 $\psi= \sqrt{({A_{max}}-{A_{min}})^2-2\sigma^2}$\\
Here $A_{max}$ and $A_{min}$ are the maximum and minimum values in the source-star DLC and $\sigma^2=\eta^2<\sigma^2_{q-s}>$, where, $Â¡Â´\sigma^2_{q-s}Â¡Âµ$ is the mean square rms error for the data points in the DLC and the error underestimation factor  ${\eta=1.54}$. The column 11 of the online Table 4 gives the mean value of $\psi$ for a session, i.e., average of the $\psi$ values estimated for the two selected DLCs of the target blazar.


\subsection{The duty cycle of variability} 
The duty cycle (DC) of intranight variability was computed according to the following definition \citep{Romero1999A&AS..135..477R}:
\begin{equation} 
DC = 100\frac{\sum_\mathbf{i=1}^\mathbf{n} N_i(1/\Delta t_i)}{\sum_\mathbf{i=1}^\mathbf{n}(1/\Delta t_i)} {\rm \%} 
\label{eq:dc} 
\end{equation}

 Here $\Delta t_i = \Delta t_{i,obs}(1+z)^{-1}$ is the intrinsic duration of the $i^{th}$ session, obtained by correcting for $z$ of the source ($\Delta t_{i,obs}$ is listed in column 4 of the online Table 4). If variability was detected in the $i^{th}$ session, $N_i$ was taken as 1, otherwise $N_i$ = 0.\par
 
  \begin{figure}

     \includegraphics[width=0.5\textwidth,height=0.16\textheight]{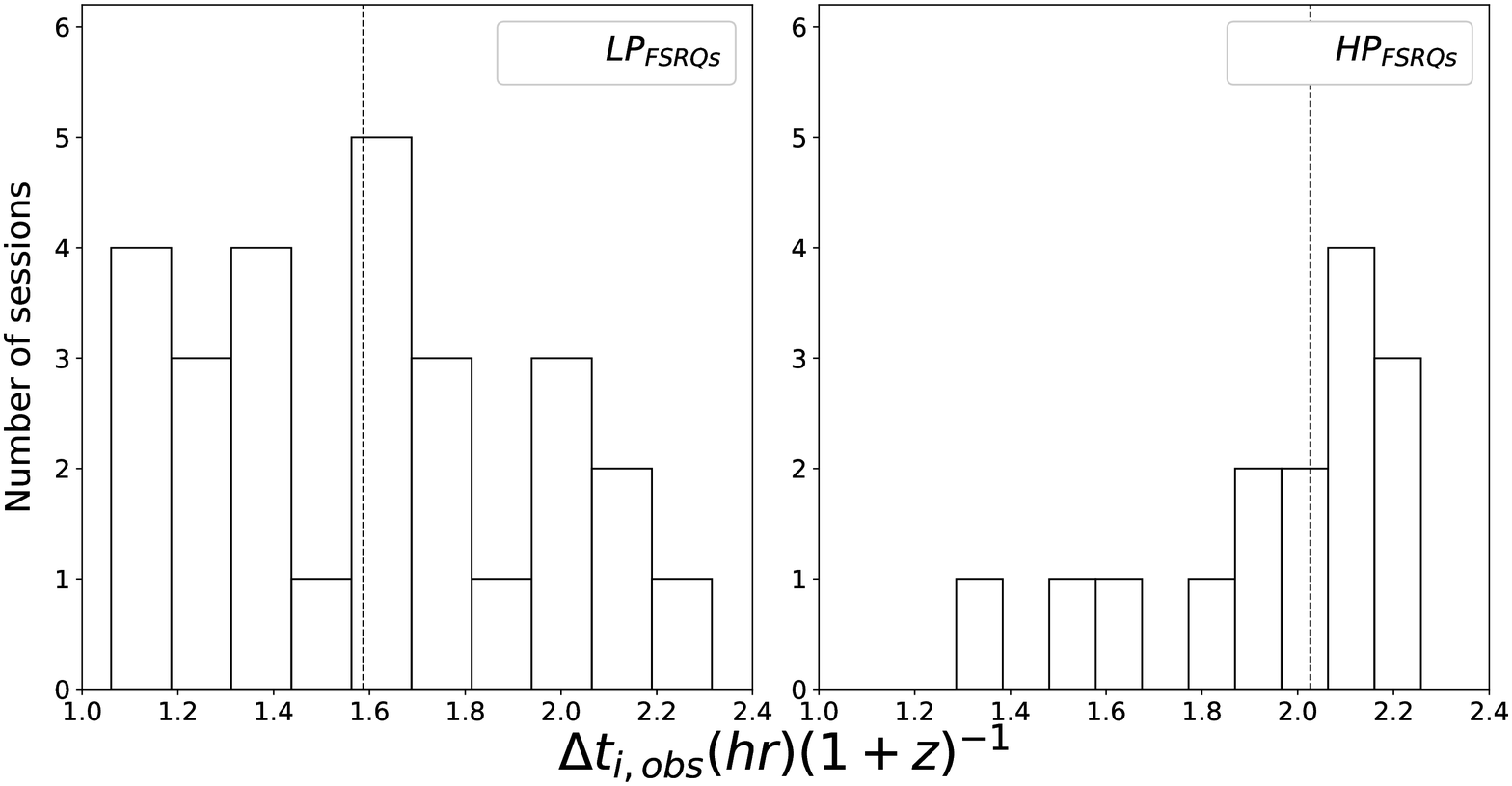}
    \vspace{-9mm}
    \caption{Histograms of the intrinsic monitoring duration ($\Delta t_{i}= \Delta t_{i,obs}(1+z)^{-1}$) for the 27 sessions devoted to the 9 $LP_{FSRQs}$ (sample-1) and the 15 sessions devoted to the 5 $HP_{FSRQs}$ (sample-2). Median of each distribution is shown with a vertical line.}
    \label{fig:all_dlc_part1} 
  \end{figure}
   
     \begin{figure}
       \includegraphics[width=0.5\textwidth,height=0.16\textheight]{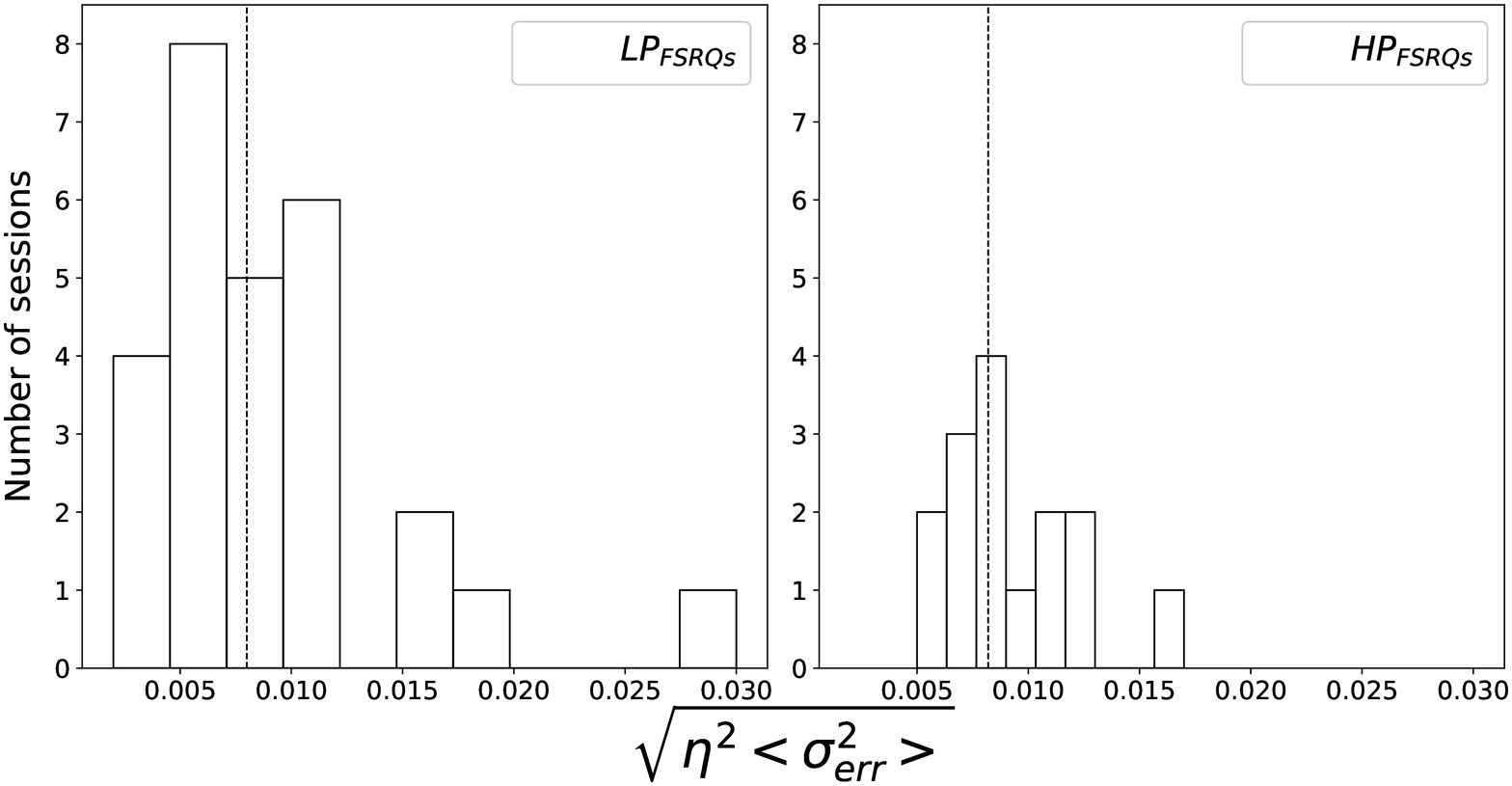}
    \vspace*{-9mm}
   \caption{Histograms of the Photometric Noise Parameter (PNP=$\sqrt {\eta^2 \langle \sigma^2_{err}\rangle}$) for the 27 sessions devoted to the 9 $LP_{FSRQs}$ (sample-1) and the 15 sessions devoted to the 5 $HP_{FSRQs}$ (sample-2). Median of each distribution is shown with a vertical line.}
 
   \label{fig:all_dlc_part2} 
  \end{figure}

\section{Results and Discussion}
\label{Discussion}

The present Red-band intranight observations of high-$z$ blazars have enabled us to monitor their rest-frame UV emission on the time scale of minutes, thus permitting this first characterisation of the intranight variability of the UV emission of blazars.
For the 9 $LP_{FSRQs}$ (sample-1), the rest-frame wavelengths of monitoring range between 1374\AA\xspace and 2078\AA\xspace (median 1973\AA\xspace), while for the 5 $HP_{FSRQs}$ (sample-2) the range is from 2078\AA\xspace to 2495\AA\xspace (median 2099\AA). The fastest time-resolved events of variability in the present campaign were recorded on 06-Apr-2020, when the DLCs of the  $LP_{FSRQs}$ J142438$+$2258 ($z = 3.62$) exhibited a sharp fading, by $\sim$ 30\% and by $\sim$ 25\%  over 20 $-$ 25 minutes, at $\sim$ 18.8 UT and $\sim$ 21.4 UT, followed by a similarly sharp recovery each time (see online Fig. 4). In the rest-frame, these durations amount to just $\sim$ 5 minutes, matching the fastest known events of blazar variability at TeV energies \citep[e.g.,][]{Aharonian2007ApJ...664L..71A,Aleksic2011ApJ...730L...8A}. 

 In the remainder of this section we shall only consider the sessions for which the variability amplitude $\psi$ is estimated to be $ > 3\%$. Now, taking only the type `V' sessions with confirmed intranight variability, the  duty cycle (DC) of the intranight variability is found to be $\sim$ 30\% for the 9 $LP_{FSRQs}$ at median $z \sim$ 2.25 (sample-1).
For the 5 $HP_{FSRQs}$ at median $z$ $\sim$ 2.0  (sample-2), the estimated DC is $\sim12\%$.
For two reasons, these estimates based on the $F_{\eta}$ test appear to be discrepant from the past INOV work where the same $F_{\eta}$ test was applied and $\eta$ was also taken to be 1.54 (see Sect.~\ref{stat}).
Firstly, using a sample of 12 moderately distant (median $z \sim 0.7$) $LP_{FSRQs}$ with $p_{opt} < 3\%$, monitored in 43 sessions of $>$ 4 hr duration, \citet{GoyalA2013MNRAS.435.1300G} estimated an INOV DC of $\sim 10\%$ for $\psi > 3\%$ cases. The DC of  $\sim {30\%}$ found here for $LP_{FSRQs}$ is much higher and, in fact, very similar to DC $\sim 38\%$ estimated by them for high-polarisation core-dominated quasars (equivalent of our $HP_{FSRQs}$). The excess of DC found here for high-$z$ $LP_{FSRQs}$ (DC $\sim$ 30\%), vis a vis the DC for their lower-$z$ counterparts (DC $\sim$ 10\%, see above) could perhaps be even greater if one were to take into account the fact that our sample of $LP_{FSRQs}$ is heavily biased towards the most luminous members of this AGN class (Sect.~\ref{The two blazar samples}) and, from several studies, variability is known to anti-correlate with luminosity. In one such study, \citet{Guo2014ApJ...792...33G} used continuum luminosity at rest-frame 2500\AA \xspace which is similar to the wavelength range covered in the present study (see above). We caution, however, that the anti-correlation with luminosity has so far been established only for variability on month/year-like time scales and it is unclear if it is significant even on the intranight time scales considered in the present study. 

One conceivable explanation for the high DC found here for sample-1 is that some of its members could be $HP_{FSRQs}$, but mis-classifed as $LP_{FSRQs}$ because their observed $p_{opt}$ (i.e., rest-frame UV polarisation) has been lowered due to dilution by the thermal UV radiation arising from the accretion disk \citep[e.g.,][]{Smith1993ApJ...415L..83S}. While such a dilution may well occur, statistical evidence for it is rather weak at present \citep[see,][]{Moore1984ApJ...279..465M,Angelakis2016MNRAS.463.3365A}. But this scenario is amenable to observational verification: (a) since the putative thermal contamination cannot alter the directional changes of the polarisation vector, one would expect to observe large directional variations for sample-1, were its members indeed $HP_{FSRQs}$; and (b) near-IR polarimetry could provide another check, since that would correspond to rest-frame optical where only a negligible suppression of the fractional polarisation due to accretion disk emission is expected. It may be noted that since any thermal dilution can only lower the fractional polarisation, any such contamination cannot result in misclassification of $LP_{FSRQs}$ as $HP_{FSRQs}$. 

Secondly, the above explanatory hypothesis for the high DC found here for high$-z$ $LP_{FSRQs}$ does not find support from the present results for the sample-2 which consists of 5 $HP_{FSRQs}$ with $p_{opt} > 3\%$ and median $z = 2.0$. For them,
 the DC of intranight variability of the UV emission is found to be quite low ($\sim 12\%$). Note that these sources cannot be $LP_{FSRQs}$ masquerading as $HP_{FSRQs}$, as noted earlier in this section. Thus, at least from the results for sample-1 and sample-2, there is no evidence for a stronger intranight variability of UV emission for high-polarisation blazars, in contrast to the blazars monitored in the rest-frame optical, for which a higher $p_{opt}$ is known to be a powerful pointer to a large intranight variability \citep{GoyalA2013MNRAS.435.1300G}.\par

It is important to check for any observational biases that might have spuriously led to the higher DC found here for the $LP_{FSRQs}$ (sample-1), in comparison to the $HP_{FSRQs}$ (sample-2). One possible factor is the rest-frame (i.e., intrinsic) duration of monitoring, since the likelihood of variability detection is known to rise for longer monitoring sessions \citep{Romero2002AA...390..431R,Carini2007AJ....133..303C}. It is seen from Fig.~\ref{fig:all_dlc_part1} that the rest-frame durations for the two samples are very similar, certainly not more for the sample-1  (median is $\sim$ 1.6 hr for the 27 sessions on sample-1 and $\sim$ 2.0 hr for the 15 sessions on sample-2). Photometric sensitivity of monitoring is another possible source of bias. Fig.~\ref{fig:all_dlc_part2} compares the distributions of the Photometric Noise Parameter PNP = {$\sqrt { \eta^2\langle \sigma^2_{i,err} \rangle}$ }, defined in Sect.~\ref{stat} (column 10 of the online Table 4). The distributions for both samples are found to have the same median PNP ( = 0.008) and therefore the lower DC found for sample-2 cannot be attributed to a more sensitive monitoring.
 
\section{Conclusions}

To sum up, if the present finding that the intranight variability of the rest-frame UV emission from $LP_{FSRQs}$ does not have a smaller duty cycle, in comparison to $HP_{FSRQs}$, were spurious, that would mean that either our two samples are too small to be representative of the high-$z$ blazar population or, alternatively, they are just a case of cosmic variance. 
An independent check on the present results would therefore be desirable and that would require intranight optical monitoring of larger, possibly independent samples of high-$z$ blazars with low and high polarisation (preferrably measured in near infrared, i.e., rest-frame optical where any polarimetric dilution due to the accretion disk emission should be a small factor). A confirmation of the present finding that between high-luminosity $LP_{FSRQs}$ and $HP_{FSRQs}$ located at high redshifts, the former exhibit a larger duty cycle of intranight variability of the rest-frame UV emission, would imply different, possibly even opposite, polarisation dependences for the intranight variability of the UV and optical radiations from blazars. The existence of such a difference would lend support to the proposal that synchrotron radiation of blazar jets in the UV/X-ray regime arises from a relativistic particle population distinct from the one responsible for their synchrotron radiation up to near-IR/optical frequencies (Sect.~\ref{introduction}). Interestingly, a hint that the medium/long-term variability of UV emission may indeed be decoupled from the optical variability, has come from spectroscopic monitoring of a large sample of 315 {\it optically-selected} high-$z$ quasars, which revealed that their observed variability is essentially confined to rest wavelengths $<$ 2500\AA\xspace (\citealp{Wilhite2005ApJ...633..638W}, also, \citealp{Punsly2016ApJ...830..104P}; \citealp{Xin2020MNRAS.495.1403X}). In view of the present findings, it is tempting to speculate that this variable UV component may be largely contributed by a jetted subset lurking among the optically-selected quasars. 


\section*{Acknowledgments}
We thank an anonymous referee for the constructive suggestions for improving the clarity of the paper.  
The assistance from the scientific and technical staff of ARIES DOT, DFOT and ST is thankfully acknowledged. G-K acknowledges a Senior Scientist fellowship from the Indian National Science Academy.
\section*{Data availability}
The data used in this study will be shared on reasonable request to the corresponding author.


\bibliographystyle{mnras}
\bibliography{references}
\label{lastpage}
\end{document}